Hyperspectral fluorescence imaging using a high-speed silicon photomultiplier array

Chi Z. Huang[1], Vincent D. Ching-Roa[1], Connor M. Heckman[1], Sherrif F. Ibrahim[2,3] and Michael G. Giacomelli[*1]

[1]Department of Biomedical Engineering, University of Rochester, 207 Goergen Hall, Box 270168, Rochester, NY 14627, USA.

[2]Rochester Dermatologic Surgery, PC, 7400 Pittsford Victor Rd Suite A, Victor, NY 14564, USA.

[3]Department of Dermatology, University of Rochester Medical Center, 601 Elmwood Ave, Rochester, NY 14620, USA

*mgiacome@ur.rochester.edu

Abstract: High-speed multiplex imaging of fluorescent probes is limited by a combination of spectral resolution, sensitivity, high cost and low light throughput of detectors, and filters. In this work, we present a hyperspectral detection system based on a silicon photomultiplier array that enables high-speed, high-light throughput hyperspectral imaging at low cost. We demonstrate 16 spectral channel imaging at 50 MP/s (800M spectra per second) with a conventional two photon microscope combined with a generalized spectral unmixing model that enables extraction of spectrally overlapping fluorophores. We show that the high spectral resolution combined with high throughput enables the multiplexing of multiple contrast agents over large areas and the detection of subtle spectral shifts associated with molecular binding. Silicon photomultiplier arrays may be a promising method to extend multiplex fluorescence imaging in a variety of scenarios.

**Introduction**

Spectral multiplexing of probes is almost universally utilized in fluorescence imaging applications such as confocal and two photon microscopy because it enables visualization of the spatial relationships between molecular species. For relatively low numbers of probes and if the emission spectra are known and distinct, individual filters can be used to divide emitted light between channels. While straightforward, this approach is difficult to scale beyond 3-4 detectors and rapidly becomes more expensive because the collection optics, filters, and detectors must be replicated for each additional channel. Furthermore, as the number of channels increases, so does the crosstalk and signal necessarily lost due to the transition bands between dichroic filters. Alternatively, a single detector can be paired with a tunable filter to precisely detect the emission of a single fluorophore. An arbitrary number of channels can be detected by tuning the filter and imaging repeatedly [1–3]. This approach potentially avoids the cost of purchasing many different filter sets and parallel detectors while being much more flexible if the exact emission spectra are not known in advance, but suffers from very poor light efficiency because all signal outside of the tunable filter passband is typically rejected. Thus, to image N spectral channels, imaging time is extended N times, and total light efficiency is reduced to 1/N. For dynamic or photosensitive samples, this approach can be limited by specimen dynamics and photobleaching. Imaging time can be reduced by incorporating dichroic filters to separate emission spectra into parallel bands with separate tunable filters [3], However, this approach increases the overall system complexity and cost and still forces some light throughput penalty.

A potentially more efficient approach is to build a scanning spectrometer into a confocal or multiphoton microscope such that emitted photons are continuously distributed by wavelength and detected in parallel [4,5]. Because the entire spectrum of emitted light can be recorded in a single pass, this approach can potentially resolve a large number of contrast agents without requiring repeated imaging or excessive photobleaching. Furthermore, multiple sets of costly dichroic filters are not required. Thus, if the scanning spectrometer can be made inexpensively, the overall cost of the system can be reduced even as the number of spectral channels is greatly increased. Due to these advantages, this approach has been commonly commercialized [6].

In spite of these potential advantages, existing detectors for spectrally-resolved laser scanning microscopes have significant limitations beyond cost. Many systems have used 2D CMOS or 2D CCD arrays [4,7] which have high quantum efficiency but are typically not single-photon sensitive and have low frame rates, necessitating either slow imaging rates or complex spatial-spectral multiplexing [8]. Alternatively, other works have used multi-anode photomultiplier tubes (MAPMT) to construct 1D array detectors [6,9]. However, these MAPMT arrays are extremely costly with limited lifetimes and suffer from low saturation powers, limiting imaging speed and potentially shot noise-limited SNR. Finally, hyperspectral single photon avalanche diode (SPAD) arrays have been demonstrated [10], but these suffer from low geometric fill factor, which limits sensitivity because most photons miss the active detector area. In this work, we present a new hyperspectral detection system based on a low-cost silicon photomultiplier (SiPM) array with a total module cost under

$1000. In contrast to MAPMT and SPAD arrays, these detectors have very high dynamic range and superior photon detection efficiency to conventional PMTs, enabling very high imaging rates and good sensitivity. We demonstrate high signal to noise ratio imaging at up to 50 MP/s (800 million spectra per second). We show that at these imaging rates fluorescent labels can be readily multiplexed. We also introduce a straightforward hyperspectral unmixing algorithm that can resolve molecular information from subtle spectral shifts associated with molecular binding.

**Method**

**SiPM linear array design**

The SiPM linear array is composed of 16 individual Hamamatsu S14160 detector channels each with an active area of 1.0x1.0 mm and an SPAD cell size of 15 microns on a single carrier PCB. Individual SiPM elements have a 0.2 mm dead space between each active area, giving an array fill factor of 83%. Each SiPM channel has 4,356 SPADs and a saturation detection rate of greater than 50 billion photons per second (>800 billion/s for the array). The PDE varies from 40% at peak (460nm) to approximately 18% at 700nm with an excess noise factor of nearly 1.0. While the PDE approximately halves at the red end of the spectrum, the extremely low excess noise factor and higher peak PDE leads to a higher sensitivity than a bialkali PMT over the entire visible spectrum. Each SiPM in the array is connected to a pole zero cancellation (PZC) filter and then amplified by a transimpedance amplifier (TIA) using the OpenSiPM design [11]. By placing the PZC before the TIA, the bandwidth and dynamic range of the circuit are dramatically extended because the amplifier sees only the fast component of the SiPM impulse response [11]. After transimpedance amplification, 16 external 27 MHz low pass filters were used to limit bandwidth to below the Nyquist rate for the A/D Individual SiPMs were biased at 49V (11V above breakdown) using a single high power boost converter developed by the OpenSiPM project [11] and powered over USB-C. The SiPM linear aaray design is available on GitHub [12]. Digitization was performed by a 16 channel A/D converter (ATS9416, Alazar Technologies) operating at 79.9 MHz, and sampled data was linearized in real-time [13].

**Microscope design**

The hyperspectral detection module was retrofitted onto an existing two photon microscope  described previously [13]. Scanning is performed by a 12kHz resonant scanner module (LSK-GR12, Thorlabs Inc.), enabling 50 MP/s video-rate imaging, while mosaic imaging uses synchronized stage position sampling to perform mosaicking at nearly 100% duty cycle. To enable hyperspectral imaging, the primary dichroic was moved from behind the objective back aperture to before the scanner entrance pupil, as shown in Fig 2, for descanned detection. A 1.33X beam expander (AC254-075-A, AC254-100-A, Thorlabs, Inc.) expands the emission to fill a high throughput diffraction grating (Ibsen Photonics FSTG-VIS1379-911). A pair of 1-inch achromatic lenses (AC254-075, Thorlabs, Inc.) focus the diffracted emission light onto the linear SiPM array. The entire spectrometer and detector cost less than $1000 excluding the A/D converter.

**Spectral calibration**

Spectral calibration was performed by comparing the SiPM measurements with those obtained from a commercial spectrometer (Ocean Optics USB2000+), using the emission spectrum of the same white LED as the reference. The SiPM hyperspectral module recorded LED spectrum was quadratically interpolated to compensate for axial color and detector rotation during assembly and then compared to the reference. The spectral range of the SiPM module was measured to be 426nm to 710nm.

**Light transmission measurement**

A 635nm laser beam was sent into the system from the pupil plane. The transmitted laser power was measured before entering the tube lens (TL) and after key components of the microscope using a power meter (Newport 843-R).

**Fluorescent Bead Phantom Imaging**

Green, yellow orange, and red fluorescent chalk dusts were dissolved in 0.5 mL of water. 1 mL of curing agent (Sylgard 182) was added to the solution and then sonicated for 20 minutes to ensure thorough mixing. 9 mL of Sylgard polydimethylsiloxane (PDMS) was added and sonicated for 10 additional minutes. The final mixture was placed under vacuum to remove bubble formation and then cured overnight. Finally, the resulting phantom was imaged using a 10X objective (Nikon CFI Plan Lambda D 10x/0.45NA) with 1040 nm two photon excitation.

**Mouse Tendon Imaging**

Formalin-fixed paraffin-embedded (FFPE) mouse tendon slides were deparaffinized, rehydrated, and stained with several different protocols.

Safranin O (SO) staining protocol: The hydrated slide was rinsed in 1% acetic acid for 15 seconds, stained in 0.1% (100ug/ml) aqueous SO solution for 5 minutes, and rinsed under running water for 3 minutes. The slide was then mounted with mounting medium with #1.5 cover glass and imaged using a 20X objective (Olympus LUCPlanFL N 20x).

Safranin O (SO), Sulforhodamine 101 (SR101), SYBR Green (SGr) staining protocol: the hydrated slide was rinsed in 1% acetic acid for 15 seconds, stained in 0.1% (100ug/ml) aqueous SO solution for 5 minutes, and rinsed under running water for 3 minutes. The slide was stained with 0.01ug/ml SR101 and 10X (0.1% stock concentration) SGr in 70% EtOH. Finally, the slide wass mounted with mounting medium with #1.5 cover glass, and imaged using a 16X water immersion objective (Nikon N16XLWD-PF-16X).

**Human Tissue Collection and Imaging**

Human surgical margins discarded after cryosectioning for Mohs surgery were collected under a protocol approved by the Research Subjects Review Board (RSRB), which serves as the IRB for the University of Rochester. Samples were cryosectioned into 50 µm thick sections and immediately placed in 95% EtOH for storage. The dehydrated tissue sections were stained in 1 mg/ml Phloxine B (PhB), and 40X SYBR Green (0.4% stock concentration, SGr) in 20% ethanol for 3 minutes. These were rinsed in water for 1 minute, mounted by mechanically pressing against to a #1.5 cover glass in a custom jig, and imaged using the 16X water immersion objective.

**Result**

**Transmission measurement**

The overall transmission of the hyperspectral system varied between 59.4% and 67.4% due to the variable diffraction efficiency of the grating, with a minimum measured at longer wavelengths and a maximum at shorter wavelengths. In addition to the grating, the other major source of loss was the 83.3% fill factor of the SiPM array. The overall measured light collection efficiency of the microscope was 28.4% due to the very poor visible throughput of the microscope scanners, which were designed around non-descanned collection and thus not coated for visible light. Thus, while the overall efficiency of the hyperspectral module was high, sensitivity was somewhat limited by the retrofitted microscope system. This could however be addressed by using a microscope designed for visible transmission and is not a limitation of the detection itself.

Components Transmission

**Generalizing PICASSO Unmixing to Hyperspectral Data**

The Process of ultra-multiplexed Imaging of biomolecules via the unmixing of the signals of spectrally overlapping fluorophores (PICASSO) algorithm performs iterative subtraction between spectral channels to suppress crosstalk by minimizing mutual information [14]. With the addition of selective input mosaicking regions through structural similarity scores (SSIM) for PICASSO (Mosaic PICASSO), Mosaic-PICASSO provides improved unmixing without overcorrection [15]. Both PICASSO and Mosaic-PICASSO has been shown to provide better unmixing results compared to conventional linear unmixing which tend to under-correct for crosstalk. This is partly due to linear unmixing relying on rigid spectral information that change within the tissue due to various binding mechanisms. Due to developed for traditional collections schemes using dichroic filters to separate collection channels, the inputs to PICASSO are assumed to be one channel per contrast agent, as would be produced by a system using fluorescent filters chosen to approximately match the emission spectra of each fluorophore. As a result, the algorithm can only subtract crosstalk but cannot combine signals split across many hyperspectral channels. A naive approach is to sum hyperspectral channels to concentrate as much of the signal of interest into a single channel, but this discards information about the shape of the spectrum and consequently struggles to unmix fluorophores with identical emission bands but different spectral shapes. To address this, we generalized the existing Mosaic-PICASSO [15] to hyperspectral image datasets (hyperPICASSO) (Fig. 3) by adding an additional data dimension reduction and unmixing step. In this algorithm, we identify approximate spectra of interest from the dataset, perform approximate unmixing to concentrate the signal into a single channel, and then perform PICASSO unmixing to fully separate the contrast channels to eliminate residual crosstalk as measured by mutual information.

**Fluorescent Bead Imaging**

We first tested the hyperspectral detector and PICASSO unmixing method [14] on fluorescent beads (Fig. 4). The fluorescent spectrum of each color bead is plotted based on the hyperspectral pixel from circled fluorescent bead pixels. Four different fluorescent beads have different peaks around 530nm, 580nm, 620nm, and 670nm with wide emission bands (Fig. 4b). As expected, these could be easily separated.

Hyperspectral imaging of highly spatially and spectrally overlapping fluorophores

Hyperspectral detection enables discrimination when using fluorophores with nearly identical emission maxima emitting from the same location. To explore this, we labeled human skin tissue with the DNA stain SYBR Gold (emission peak: 537 nm), which exclusively labels cell nuclei, and the protein counterstain Eosin-Y (emission peak: 545 nm), which labels all proteins, including those in nuclei. As a result, the emission spectra (Fig. 5e) of cell nuclei and stroma are extremely similar, with only minor differences in the shape of the spectra. Accordingly, summing spectral bins and then attempting conventional Mosaic-PICASSO unmixing struggles to separate cell nuclei from stroma (Fig. 5 a,c), for any combination of spectral bins selected. Conversely, our generalized hyperPICASSO algorithm incorporates information about the overall shape of the emission spectrum and more easily separates the stroma channel from the nuclei channel (Fig. 5 b,d). The resulting two channel images show substantial spectral leakage in the conventional Mosaic-PICASSO case with binning (Fig. 5f) but strongly separated stroma and nuclei channels with hyperPICASSO (Fig. 5g).

**Molecular discrimination based on spectral shifts in human surgical specimens**

We next explored if hyperspectral imaging could reveal molecular information about the composition of human tumors that could be used to help diagnose disease or guide surgical excision. We selected discarded squamous cell carcinoma (SCC) samples from patients treated at our institution and labeled them with SYBR Green (SGr, a DNA stain) and Phloxine B (PhB, an Eosin analog). PhB is a histology stain that labels most proteins but has a distinct hue upon binding to keratin [17,18], a common component of skin that is often present in epithelial tumors such as SCC. We hypothesized that the redder hue observed upon binding to keratin in conventional histology would result in a distinctive fluorescent emission spectrum upon binding to keratin. Accordingly, imaging of SCC samples (Fig. 6) revealed two distinctive emission spectra, one for general protein and another, green-shifted spectrum for keratin-bound PhB. From these two stains and using hyperPICASSO, we extracted 4 contrast channels, DNA, keratin, collagen, and stroma (Fig. 6a,b). From this data, we generated virtual trichrome stained transmission images (Fig. 6c), labeling keratin blue-green. As expected, these distinctively label both the keratin pearls through the SCC tumor core as well as the normal keratinized epithelium while excluding the normal non-keratinized dermis and underlying connective tissues with high contrast (Fig. 6d).

## Spectral Multiplexing of overlapping dyes

Hyperspectral imaging also enables multiplexing multiple fluorescent labeling with similar emission bands that are ordinarily difficult to separate. Fig. 7 shows multiplexing of Sulforhodamine 101 (SR101) with an emission peak at 595nm, Safranin O (SO) peak at 575nm, and SYBR Green (SGr) with the mission peak at 525nm staining on the mouse knee join slide. Using hyperPICASSO unmixing, muscle/spongy bone region (Fig. 7a), knee growth plate stem cells (Fig. 7b), compact bone region (Fig. 7c), nuclei (Fig. 7d), and collagen with SHG (Fig. 7e) channels can be separated in spite of all emission peaks fitting into ~ 70 nm of band of spectral range. Notably, the SO emission spectrum primarily labels the stroma which generally overlaps with SR101 staining except that SO preferentially stains muscle and spongy bone, while SR101 preferentially stains compact bone and tendon (Fig. 7f). SO bound to the growth plate also shows a distinctive spectrum which can be easily separated. while areas where SHG is disrupted by out of plane folding of fibers are readily identified with strong SR101 staining, but without SHG signal. Conversely, we found that the standard Mosaic-PICASSO algorithm could not reliably separate the SHG channel from the overlapping DNA channel without access to the shape of the SGr emission spectrum.

## Discussion

We present a hyperspectral detection system using 16 parallel transimpedance amplifiers connected to 16 high dynamic range, single photon sensitive silicon photomultipliers. In contrast to alternative detection technologies such as CMOS arrays, the analog bandwidth of each array element is 75 MHz, fast enough to enable read out of the fastest point scanning systems. In contrast to multi-anode PMTs, each SiPM element is completely independent with zero inter-channel crosstalk, and a saturation power of approximately 40 billion photons per second. As the overall array throughput approaches 1 trillion photons per second, saturation from fluorescence is virtually nonexistent leading to extremely high linearity. In comparison to other types of SPADs, the read out electronics of a SiPM are external to the array, enabling relatively high fill factors and photon detection efficiencies comparable to or exceeding that of conventional photomultiplier tubes. In addition, the overall costs of both the external electronics and photodetectors are extremely low, with individual SiPM and read out electronics costing less than $20 per channel in low volumes and substantially less if manufactured in volume. This cost is dramatically lower than competing vacuum tube photomultipliers or high-end CMOS or EMCCDs. Thus this technology could be readily incorporated into a variety of fluorescence instruments for minimal added cost.

A further advantage is the extremely high dynamic range, negligible crosstalk and near-perfect linearity of SiPM detectors at typical fluorescence power levels. As a result, each additional fluorophore adds linearly to the existing channels, greatly simplifying computational unmixing of fluorescent species. We utilize this ability to demonstrate four-way discrimination of highly overlapping spectral labels that of a green fluorescent DNA label (emission peak: 525nm), SHG signal (emission peak: 520nm), and two different green fluorescent spectra from phloxine B (peak:580 nm) based on its binding to keratin with high SNR and very high imaging rate. Similarly, SYBR Gold (emission peak: 537nm) and Eosin-Y (emission peak: 545nm) could be readily separated in spite of near-total spectral overlap. This approach could be readily extended to multiplexing of antibodies, DNA-based probes, or fluorescent proteins enabling very dense molecular labeling.

To utilize the high linearity and photon throughput of the SiPM array, we generalized the PICASSO algorithm to hyperspectral datasets by adding a straightforward data dimension reduction step that concentrates information spread across the emission spectra into a single channel which then can be unmixed using mutual information relative to other channels. We found that as compared to naive data dimension reduction strategies, such as binning the spectra bins with the majority of each fluorophore's emission, that this enabled dramatically better extraction of features because it incorporates information about the shape of the emission spectra and not simply the center wavelength.

The ability to visualize keratin within minutes of excision is intriguing in the context of histopathology of cancer, as labeling of keratin labeling is challenging with conventional stains [20]. While immunohistochemistry can be used, this is a time-consuming process that is impractical to perform intraoperatively due to long delays. Thus, no method exists for assessment of keratinized carcinomas during treatment of either skin or head and neck cancers, two extremely types of cancer that impose a large treatment burden [21,22]. We demonstrate here that hyperspectral imaging combined with spectral unmixing can directly discriminate keratinized structures in human squamous cell carcinomas with a simple stain that can be rapidly applied to fresh tissue. In comparison, frozen section analysis of soft tissues such as skin can take tens of minutes for preparation, while sectioning of bony tissues as in many head and neck cancers is an extremely time-consuming process that can greatly extend surgery. The high imaging rate demonstrated here enabled imaging entire excision surfaces in less than the freezing time for cryosectioning.

The main limitation of the presented system was the overall light transmission of the microscope combined with the additional losses from the hyperspectral module. While the microscope we retrofitted had a high loss in the visible spectrum, the transmission of the actual hyperspectral module was relatively high at approximately 60 to 70%, depending on wavelength. While low compared to the very high throughput of typical dichroic filters, when multiplexing probes with spectral unmixing the entire emitted spectrum can be utilized, rather than just the fraction that falls within the passband of the dichroic. Furthermore, while the 73-90% diffraction efficiency of the grating is hard to improve further over such a wide spectrum, a newer generation of SiPM arrays has become available that improves the array fill factor from 83% to 94% while improving quantum efficiency [23]. If utilized here, the improved detector model would increase overall efficiency at the peak wavelength to nearly 80%. It may also be possible to adopt this approach to nondescanned detection utilizing larger array elements, which would enable deep tissue hyperspectral imaging. Thus, the relatively low transmission of the available microscope was not an intrinsic limitation of the technique and could be readily improved. Another disadvantage is the need for a 16 channel ADC on the microscope, but if spectral unmixing or very high imaging rates were not required it, the high gain of the SiPM array would make it straightforward to subsample or multiplex onto a 4 channel ADC as commonly used on microscopes.

**Conclusion**

We present a novel hyperspectral detector based on a SiPM array combined with a hyperspectral unmixing algorithm that enables rapid (800 Mspectra/s) imaging of large tissue samples and then extraction of distinct molecular labels. We demonstrate that fluorophores with near-complete spectral and spatial overlap can be readily discriminated based on the emission spectral shape and that subtle shifts in the fluorescent emission of labels can be used to identify what the label is bound to. This approach may have applications to histopathology, multiplex fluorescent imaging and other applications.

Funding. This study was supported by the US National Institutes of Health, Grant No. R37-CA258376 and R21-EB032839.

Acknowledgments. We thank Beth Geer for helping to collect discarded human skin tissue. We thank Dr. Whasil Lee for providing formalin-fixed paraffin-embedded (FFPE) mouse tendon slides.

Disclosures. The authors declare no conflicts of interest.

Data availability. The dataset used and/or analyzed during the study are available from the corresponding author on reasonable request.

**Figures and Tables**

Table 1. Overall transmission

| Components | Transmission |
|---|---|
| Scanner, scan lens, objective, dichroic, and tube lens throughput | 47.80% |
| Hyperspectral module throughput | 59.4%-67.4% |
| Total collection throughput | 28.50% |

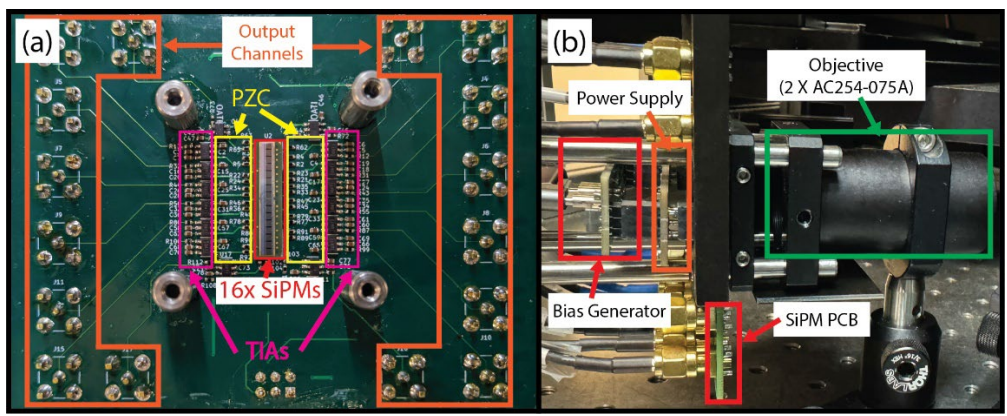

Fig. 1. OpenSiPM-based linear SiPM array and power supply. (a): Detection side (facing grating) of linear SiPM array board with 16 element SiPM array, pole zero cancellation (PZC) filters, and then transimpedance amplifiers (TIAs). (b): Side view of the OpenSiPM-based hyperspectral detection model with bias generator, power supply, and the main SiPM PCB. The objective lens contains two stock Thorlabs achromatic lenses (AC254-075A).

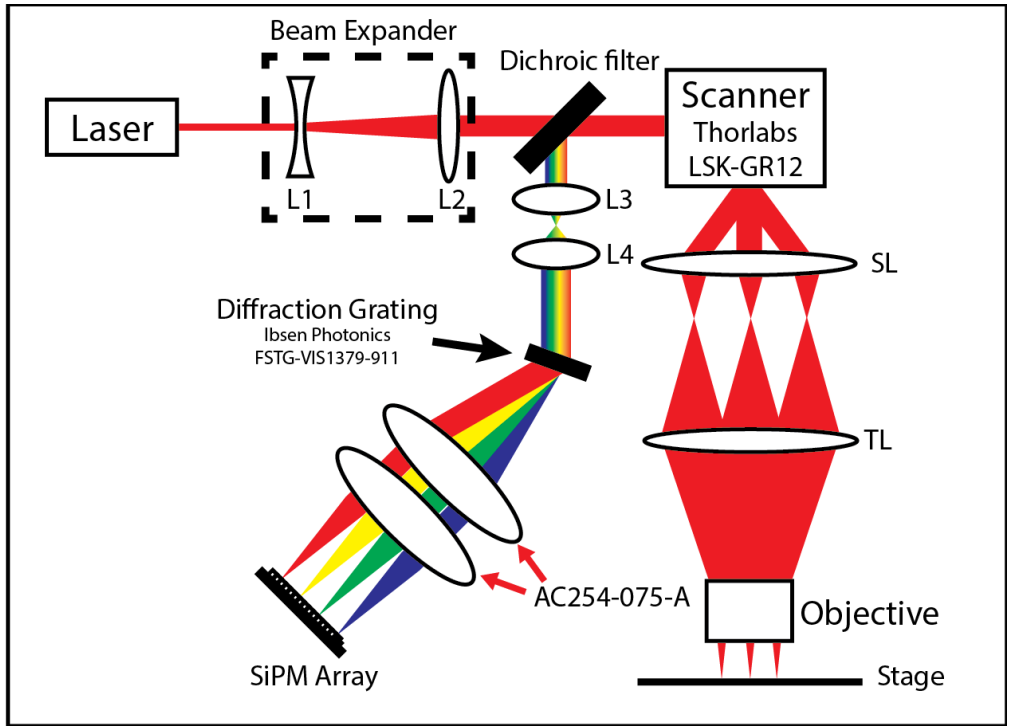

Fig. 2. Optical design of the descanned the hyperspectral two photon microscope. L1: Thorlabs LC1582-B (f=-75mm), L2:Thorlabs LA4924-B (f=175mm), L3:Thorlabs AC254-075-A (f=75mm), L4: Thorlabs AC254-100-A(f=100mm), SL: Thorlabs SL50-CLS2, TL:TTL200MP.

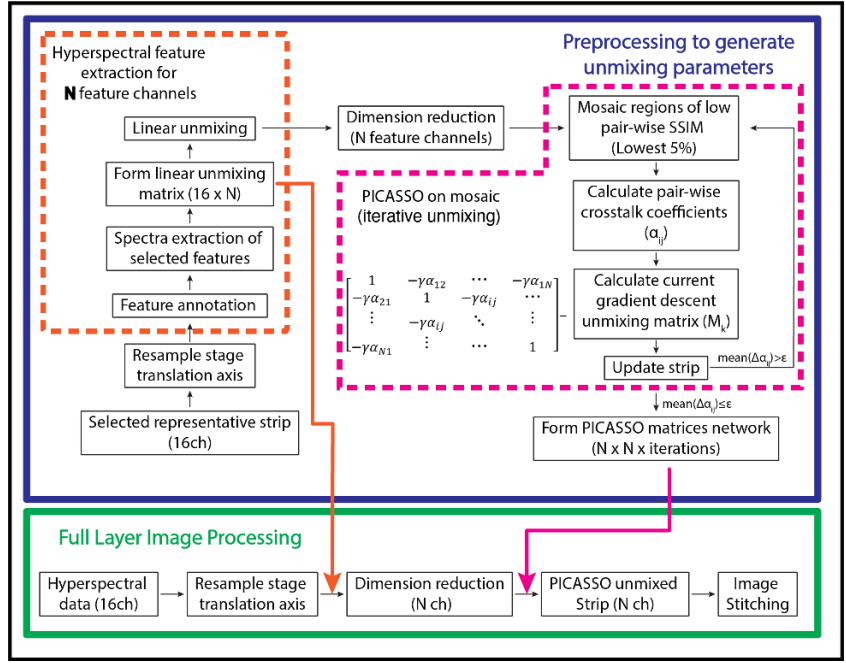

Fig. 3. Generalized hyperspectral PICASSO (hyperPICASSO) unmixing. A preprocessing step (purple box) is used to extract features of interest from a hyperspectral dataset and then develop a set of iterative unmixing matrices. Following processing, these can be performed per pixel on hyperspectral data, either offline or in real-time as new data is acquired (green box). Furthermore, we combine this with synchronous pixel sampling [16] to enable rapid assembly of gigavoxel-scale mosaic images.

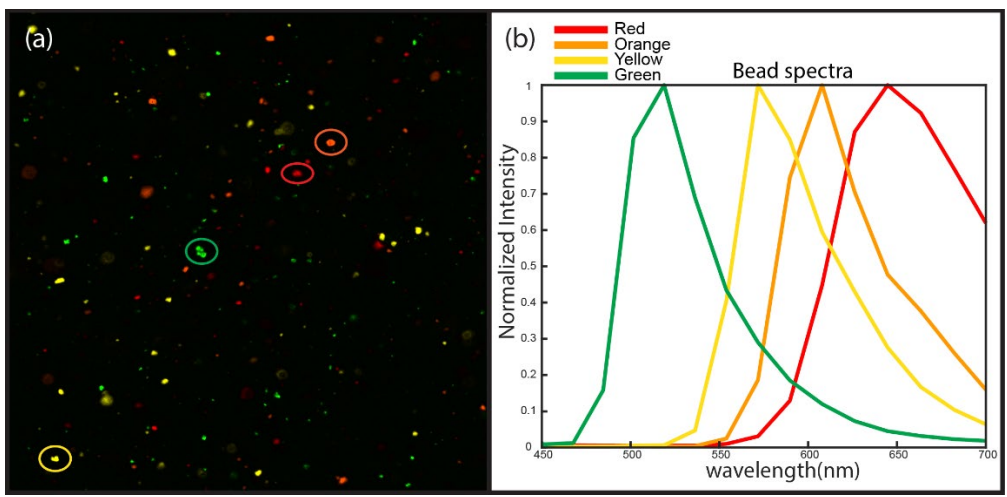

Fig. 4: Hyperspectral fluorescent bead imaging. (a): Hyperspectral image of four different color fluorescent particles suspended in Polydimethylsiloxane (PDMS). (b): emission spectrum of each color in corresponding circles.

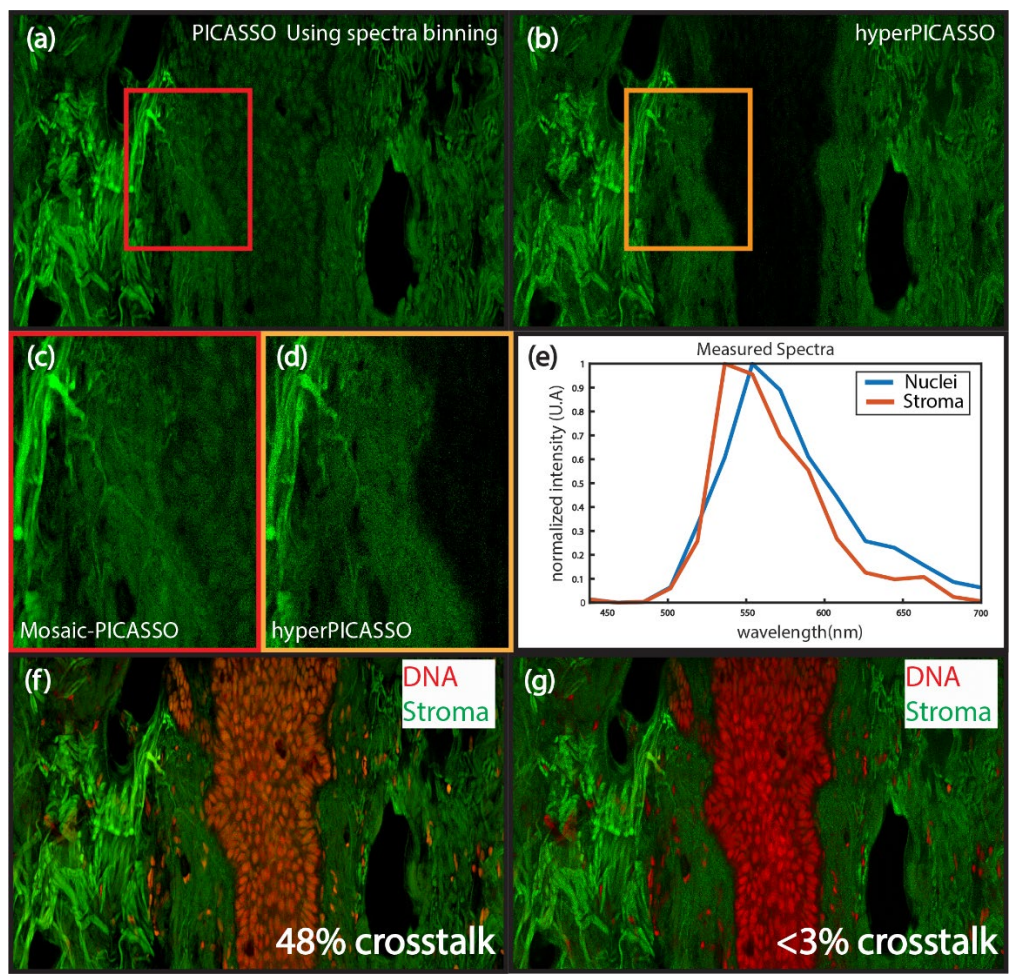

Fig. 5. Mosaic-PICASSO unmixing with spectra channel binning vs linear unmixing with measured emission spectra. (a): Unmixed stroma channel image using Mosaic-PICASSO unmixing method with spectra binning. (b): Unmixed stroma channel image hyperPICASSO. (c): Boxed region in (a) showing residual nuclei signal remaining in the Mosaic-PICASSO only method. (d): Boxed region in (b) and same location as (c) showing accurate unmixed stroma channel without significant residual nuclei signal. (e): Measured nuclei and stroma fluorescent emission spectra from Sybr Gold and eosin-stained human skin tissue. (f): Combined unmixed DNA (red) and Stroma (Green) channel using the Mosaic-PICASSO method in (a). (g): Combined unmixed

DNA (red) and Stroma (Green) channel using the hyperPICASSO showing reduction in crosstalk of the stroma from 48% to less than 3%.

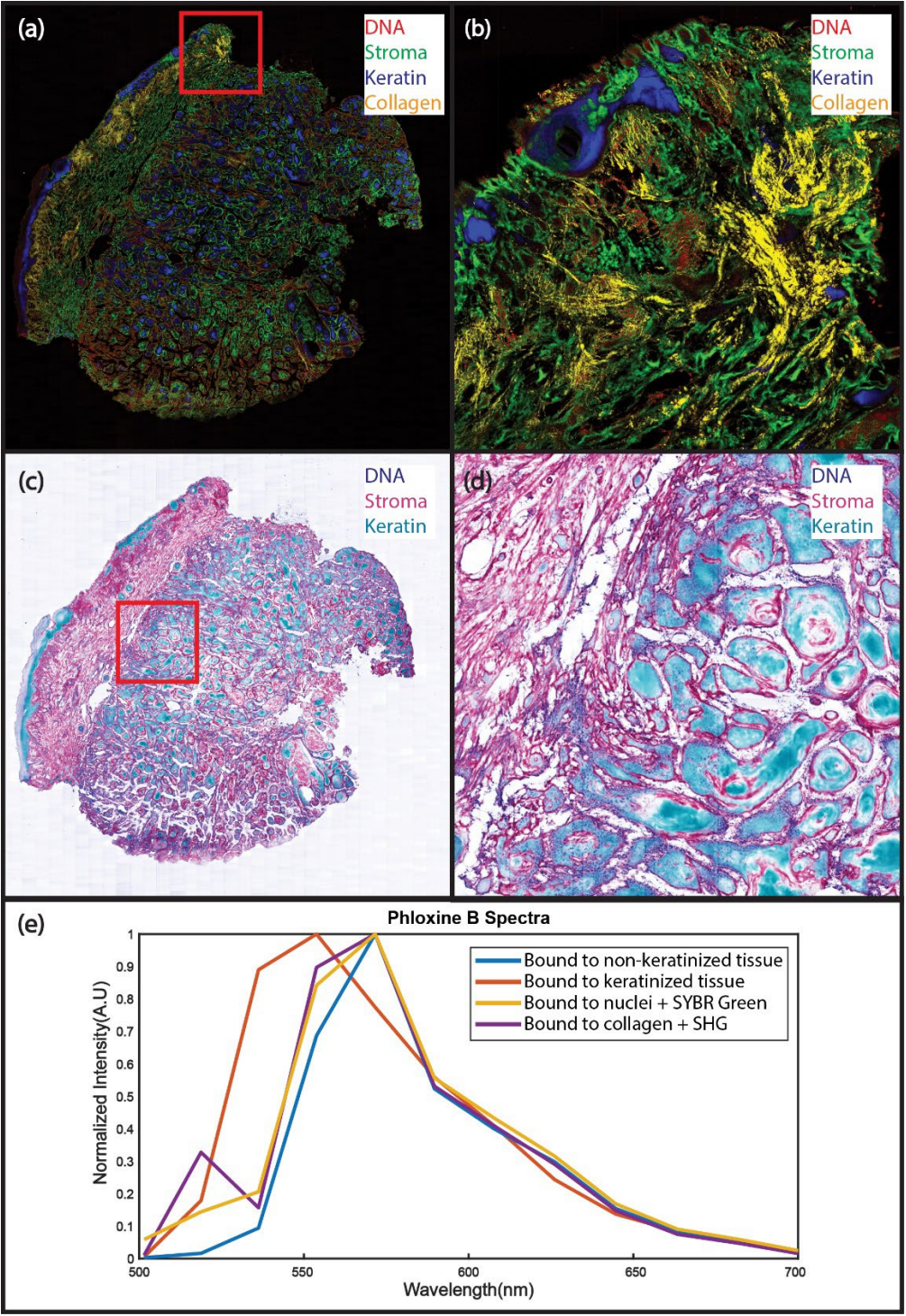

Fig. 6. Human surgical margin positive for squamous cell carcinoma labeled with SYBR Green (DNA) and Phloxine B (protein and keratin) imaged at 50 MP/s with hyperPICASSO unmixing. (a): Spectrally unmixed mosaic image with DNA red, the non-keratin bound PhB green, keratin-bound PhB blue, and collagen Yellow. Full resolution image link: https://imstore.circ.rochester.edu/hyperspectral/PhBSGrSHG/zstackRgb.html (b): The enlarged region in B shows keratin-rich regions, such as the epidermis and hair follicles, as well as hallmarks of squamous cell carcinoma, such as keratin pearls. Strong SHG single is from dense connective tissue forming due to a previous biopsy. (c): The mosaic in (a) is rendered as a virtual trichrome stain using the virtual transillumination algorithm [19] Full resolution image link: https://imstore.circ.rochester.edu/hyperspectral/PhBSGrTri/zstackRgb.html (d): The enlarged region shows PhB bound to different tissue types. (e): Spectra from regions in A - D showing the change of PhB emission upon binding to keratinized tissue(orange vs blue), combining emission with SGr on nuclei with additional emission in green (yellow vs. blue), and combining emission on collagens with additional SHG peak(purple vs. blue). Imaging time for the 220 mm2 slide was less than 2 minutes.

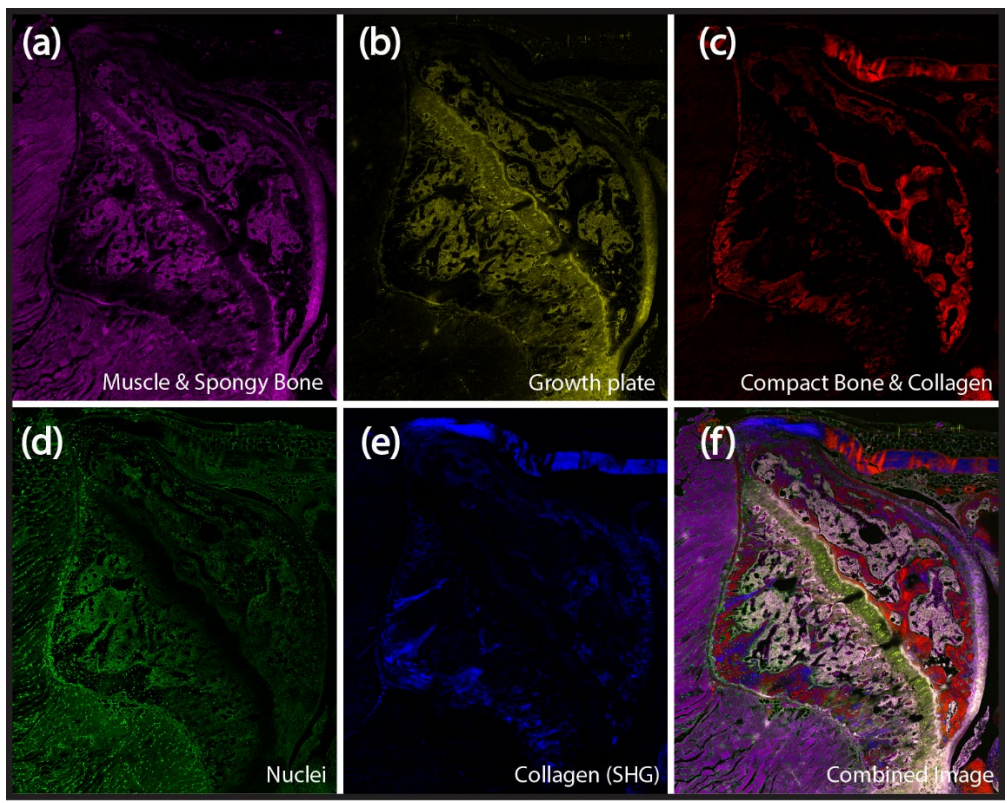

Fig. 7. Hyperspectral image of mouse tendon with Sulforhodamine 101 (SR101), Safranin O (SO), and SYBR Green (SGr) staining excited at 1040nm unmixed using hyperPICASSO. (a): Unmixed muscle and spongy bone image, predominately fluorescent emission from muscle-bound SO with slight emission green shift combined with SR101 emission. (b): Growth plate and spongy bone channel, predominantly stained by SO with a slight emission red shift. This region is similar to the region of SO stains in traditional histology. (c): Unmixed compact bone and collagen image, predominantly from SR101 staining with weak SO staining. (d): Nuclei channel from SGr emission. (e): Collagen channel from SHG showing tendon, cartilage, and bone. (f): The combined image of muscle, bone, growth plate, nuclei, and collagen channels in a-e. The total image time for the 131 mm2 slide was 40 seconds. Full resolution image link: https://imstore.circ.rochester.edu/hyperspectral/Mousejoin_SRSOSGr_16X/zstackRgb.html